\documentclass{llncs}
\usepackage{llncsdoc}
\usepackage{graphics}
\usepackage{graphicx}
\usepackage[export]{adjustbox}
\usepackage{amsmath,amsfonts,latexsym,amssymb,euscript,xr}

\usepackage{color}

\begin{document}

\title{Constraint-based modeling and simulation of cell populations}
 
\author{Marzia Di Filippo\inst{1,4,\dagger} \and Chiara Damiani\inst{1,2,\dagger,*} \and Riccardo Colombo\inst{1,2} \and \\
Dario Pescini\inst{1,3,*} \and Giancarlo Mauri}
\institute{SYSBIO Centre of Systems Biology, Piazza della Scienza 2, 20126 Milano, Italy \and Dipartimento di Informatica, Sistemistica e Comunicazione, \\
Universit\`a degli Studi di Milano-Bicocca, Viale Sarca 336, 20126 Milano, Italy \and Dipartimento di Statistica e Metodi Quantitativi, \\
Universit\`a degli Studi di Milano-Bicocca, Via Bicocca degli Arcimboldi 8, 20126 Milano, Italy \and Dipartimento di Biotecnologie e Bioscienze, \\
Universit\`{a} degli Studi di Milano-Bicocca, Piazza della Scienza 2, 20126 Milano, Italy}

\maketitle

$\dagger$ Equal contributors\\
$*$ Corresponding authors: {chiara.damiani,dario.pescini}@unimib.it

\begin{abstract}
The intratumor heterogeneity has been recognized to characterize cancer cells impairing the efficacy of cancer treatments. We here propose an extension of constraint-based modeling approach in order to simulate metabolism of cell populations with the aim to provide a more complete characterization of these systems, especially focusing on the relationships among their components. We tested our methodology by using a toy-model and taking into account the main metabolic pathways involved in cancer metabolic rewiring. This toy-model is used as ``individual'' to construct a ``population model'' characterized by multiple interacting individuals, all having the same topology and stoichiometry, and sharing the same nutrients supply. We observed that, in our population, cancer cells cooperate with each other to reach a common objective, but without necessarily having the same metabolic traits. We also noticed that the heterogeneity emerging from the population model is due to the mismatch between the objective of the individual members and the objective of the entire population.

\end{abstract}

\section{Introduction} \label{sec:intro}
A reprogramming of cellular energy metabolism has recently been included within the hallmarks \cite{hanahan2011} of cancer. An overall rewiring of metabolism is indeed fundamental to most effectively support the uncontrolled and enhanced growth characterizing all tumor cells. \\
An attention on the single molecules that are responsible for cancer onset fails to handle the non-linearity and complexity of cancer metabolic rewiring \cite{alberghina2005}. For this reason, metabolomics aims at concurrently identifing and quantifing the full set of metabolites that are present within a given cell or tissue type at a given time, thus providing a snapshot of the cell phenotype \cite{griffin2004,lee2006,cazzaniga2014}.\\
As a matter of fact, information and knowledge can be extracted from these large collections of data only by rationalizing and integrating them into computational predictive models.
In this regard, constraint-based modeling has been by far the most applied technique to the study of metabolism. It  indeed represents the best compromise between the purely qualitative information provided by graph-theory based topological models and the mechanistic details provided by kinetic modeling, which is currently impracticable for networks on a genome-scale.
In particular, Flux Balance Analysis (FBA) - which exploits Linear Programming to identify the distribution of the metabolic flux that optimizes a metabolic objective - has extensively been applied to cancer research, as maximization of growth rate may accurately describe the objective driving cancer evolution \cite{folger2011,agren2012,agren2014,gatto2014}. 

Classic FBA is limited to the simulation of a single (or average) cell that is representative of the metabolism of the entire population this cell belongs to. This is a major drawback if we consider that a cell population is not necessarily homogeneous and various metabolic phenotypes may be generated \cite{tseng2016,styczynski2016}. 
In fact, the heterogeneity characterizing cancer disease is not limited to the one existing among individual tumour types. Indeed there are multiple sources of intratumor heterogeneity leading phenotypic differences among cells belonging to the same population. Unfortunately, many anti-cancer treatments are not able to deal with intratumor heterogeneity, drastically reducing their efficacy \cite{nowell1976,mohanty2012,gerlinger2012}. As a consequence, single-cell metabolomics techniques are currently under development as a promising strategy to unravel metabolic heterogeneity among cells belonging to the same tumor, which metabolomics hides as a result of average measurements of population behavior, by investigating singularly the role of distinct cell types within a given population. However, these kind of experiments are still at an early stage and numerous technical limitations remain to be solved \cite{bunawan2016single,zenobi2013single}.

To address the issue, we propose here an extension of constraint-based modeling to study metabolism of cell populations in order to provide a more complete characterization of these systems and to investigate relationships among their components. We assume that the heterogeneity emerging from a given cell population is due to the fact that the objectives of the individual members do not correspond to the objective of the entire population.

As a proof of principle, we test our methodology with a toy-model of cancer metabolism that has been reconstructed based on the current knowledge on the metabolic pathways most involved in cancer metabolic rewiring.

\section{Flux balance analysis and flux variability analysis}\label{sec:FBAandFVA}
Flux Balance Analysis allows to calculate the optimal flux distribution, which is the rate at which each of the $R$ reactions of a network occurs at steady state.\\
By relying on a steady state assumption, according to which concentration of each of the $M$ metabolites belonging to the network remains constant over time, FBA does not require any knowledge on enzymatic kinetic or metabolite concentrations. The application of further constraints on the system is used to reduce the number of putative flux distributions defining an allowable solution space in which any flux distribution may be equally acquired by the network. The optimization (maximization or minimization) of a specific objective function (e.g. maximization of ATP or biomass production, minimization of nutrients utilization) allows to narrow the set of feasible solutions and to identify a single optimal flux distribution.\\
Given a $M\times R$ stoichiometric matrix  S, whose element $s_{i,j}$ takes value $-\alpha_{ji}$ if the species $S_i$ is a reactant of reaction $j$, $+\alpha_{ji}$ if  species $S_i$ is a product of reaction $R_j$ and 0 otherwise - where $-\alpha_{ji}$ is the stoichiometric coefficient of reactant/product $i$ in reaction $j$ - the problem is postulated as a general Linear Programming formulation:

\begin{equation}
\begin{split}
\text{maximize~or~minimize~} Z=\sum_{i=1}^{R}w_i v_i \\
\text{subject~to~}
S \vec{v}=\vec{0},~
\vec{v}_{min} \leq \vec{v} \geq \vec{v}_{max}.
\end{split}
\label{FBA}
\end{equation}
where $w_i$ is the objective coefficient for flux $v_i$ in vector $\vec{v}$; whereas the vectors $\vec{v}_{min}$ and $\vec{v}_{max}$ specify, respectively, the lower and upper boundaries of the admitted interval of each flux $v_i$.
A negative value $v_i$ conventionally indicates flux trough the backward reaction. 
To achieve mass balance in an open system, exchange of a given nutrient A with the enviroment is defined by unbalanced reactions in the form of $A <=> \emptyset$. 
For a more comprehensive description of FBA, the reader is referred to \cite{orth2010flux}.

Frequently, although FBA only returns a single flux distribution, the constraints imposed on the system under investigation do not allow to obtain a unique solution, but may confine the solution space to a feasible set of alternative optimal flux distributions in which the same optimal flux value of the objective function may be reached through different but equally possible ways. In this context, Flux Variability Analysis (FVA) \cite{mahadevan2003} returns the range of flux variability of each reaction, i.e. the allowable minimum and the maximum fluxes by each reaction, but it does not identify all the alternative optimal solutions. This task can be performed by exploiting recursive MILP optimization, as proposed in \cite{reed2004}.

\section{A proposal for using the constraint-based approach to model cell populations}\label{sec:proposal}
Metabolic networks reconstructed starting from genome annotation are increasingly being available for different organism, spanning from micro-organisms \cite{feist2009} to human metabolism. These networks may encompass virtually all reactions that can be catalyzed by the enzymes encoded by a given genome, or only a portion of them \cite{ryu2015}.

\begin{figure}
\centering
\includegraphics[max height=10 cm,max width=13 cm]{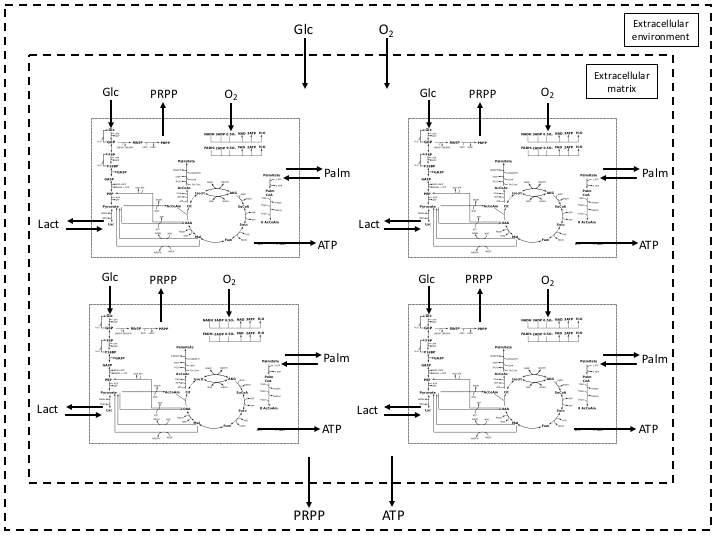}
\caption[Schematic representation of the population model]{Schematic representation of the population model. A single entity model is used as building block and replicated N times to obtain a network of metabolic networks. All the members belonging to the population model have the same topology and stoichiometry, share the same nutrients (in our case, glucose and oxygen) supply, and have the same reactions to exchange some metabolites with other components of the population (within a compartment referred to as ``Extracellular matrix''), or with the external environment (referred to as ``Extracellular environment'')}
\label{fig:1}
\end{figure}

In order to fill the existing gap between the understanding of single cells function (represented by a metabolic network) within a given tissue and their role when they are interacting with each other within a population, we propose to replicate N copies of the reference metabolic network with univocal names for metabolites and reactions, so to obtain a $(M \cdot N)\times (R \cdot N)$ stoichiometric matrix. 
For the exchange of intracellular nutrients with the environment (the extracellular matrix) of each of the $N$ networks, the unbalanced reactions $A_i <=> \emptyset$  are replaced by reactions in the form $A_i <=> A_{medium}$ where $A_i$ refers to metabolite A in network $i$, whereas $A_{medium}$ refers to metabolite $A$ in the extracellular matrix, to mimic the fact that cells in the population share the same resources.
To achieve mass balance of the population model as an open system, a set of $E$ exchange unbalanced reactions for metabolites within the extracellular matrix must be included.
Note that the set of metabolites that the cells exchange with the extracellular matrix and the set of metabolites that the cell population share with the external environment do not necessarily coincide.
A schematic representation of the population model is provided in Figure \ref{fig:1}.

Once the $(M \cdot N)\times (R \cdot N+E)$ stoichiometric matrix and the vector of objective coefficients are obtained for the population model, standard FBA can  be applied to obtain the distribution of flux across the N cells that maximize the population objective.

For this purpose, we implemented an algorithm (using the Python programming language and the COBRApy package) that automatically replicates a number of times any SBML model to obtain the above defined population model, in a suitable form to then perform FBA and FVA analyses on the generated model.

\section{Results}\label{sec:ResultPart1}
\begin{figure}
\centering
\includegraphics[max height=7 cm,max width=10 cm]{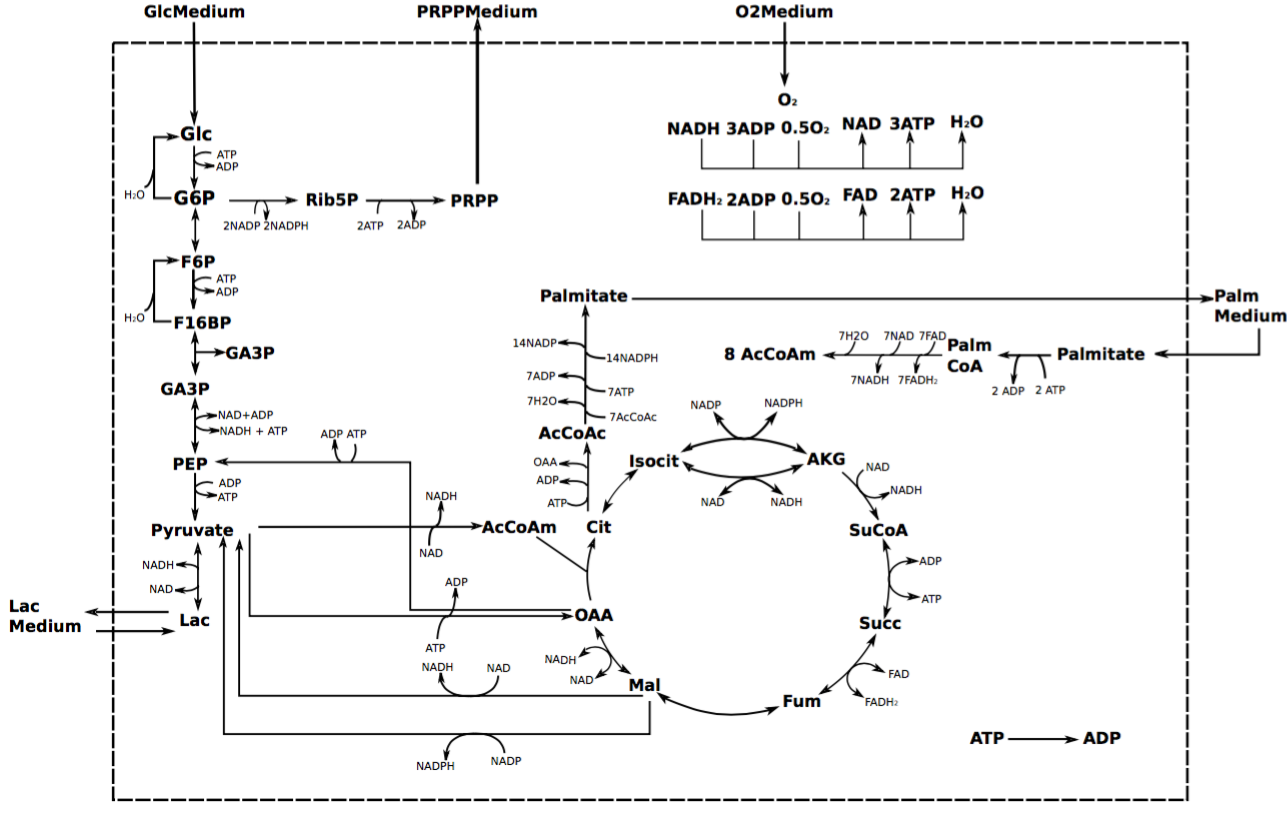}
\caption{Graphical representation of the toy-model. This map shows a representation of the toy-model used in our work. The network consists of the following metabolic pathways: glycolysis, lactate production and consumption, TCA cycle, OXPHOS, pentose phosphate pathway, palmitate synthesis and $\beta$-oxidation. Abbreviations: Glc, glucose; G6P, glucose-6-phosphate; F6P, fructose-6-phosphate; F16BP, fructose-1,6-bisphosphate; GA3P, glyceraldehyde 3-phosphate; PEP, phosphoenolpyruvate; Pyruvate, pyruvate; Lac, lactate; Rib5P, ribose-5-phosphate; PRPP, phosphoribosyl pyrophosphate; AcCoAm, mitochondrial acetyl-CoA; AcCoAc, cytosolic acetyl-CoA; Palmitate, palmitate; PalmCoA, palmitoyl-CoA; Cit, citrate; Isocit, isocitrate; AKG, $\alpha$-ketoglutarate; SuCoA, succinyl-CoA; Succ, succinate; Fum, fumarate; Mal, malate; OAA, oxaloacetate}
\label{fig:2}
\end{figure}

\begin{figure}
\centering
\includegraphics[max height=7 cm,max width=7 cm]{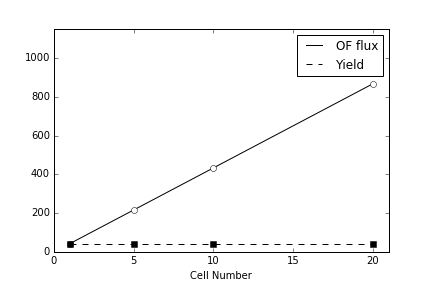}
\caption[Yield and Objective function flux values vs. Cell number.]{Variation of objective function flux value within the population model according to the increasing in the number of components of the model itself. In the graph, the continuous line shows how the flux value of the objective function of the population model (``OF flux'') increases proportionally with the increase in the number of its members (``Cell Number''). The dashed line shows how, in relation to the increase in cell number, the ratio between the OF flux and the cell number (referred to as the ``Yield'') is always the same}
\label{fig:3}
\end{figure}

\begin{figure}
\centering
\includegraphics[max height=13 cm,max width=10 cm]{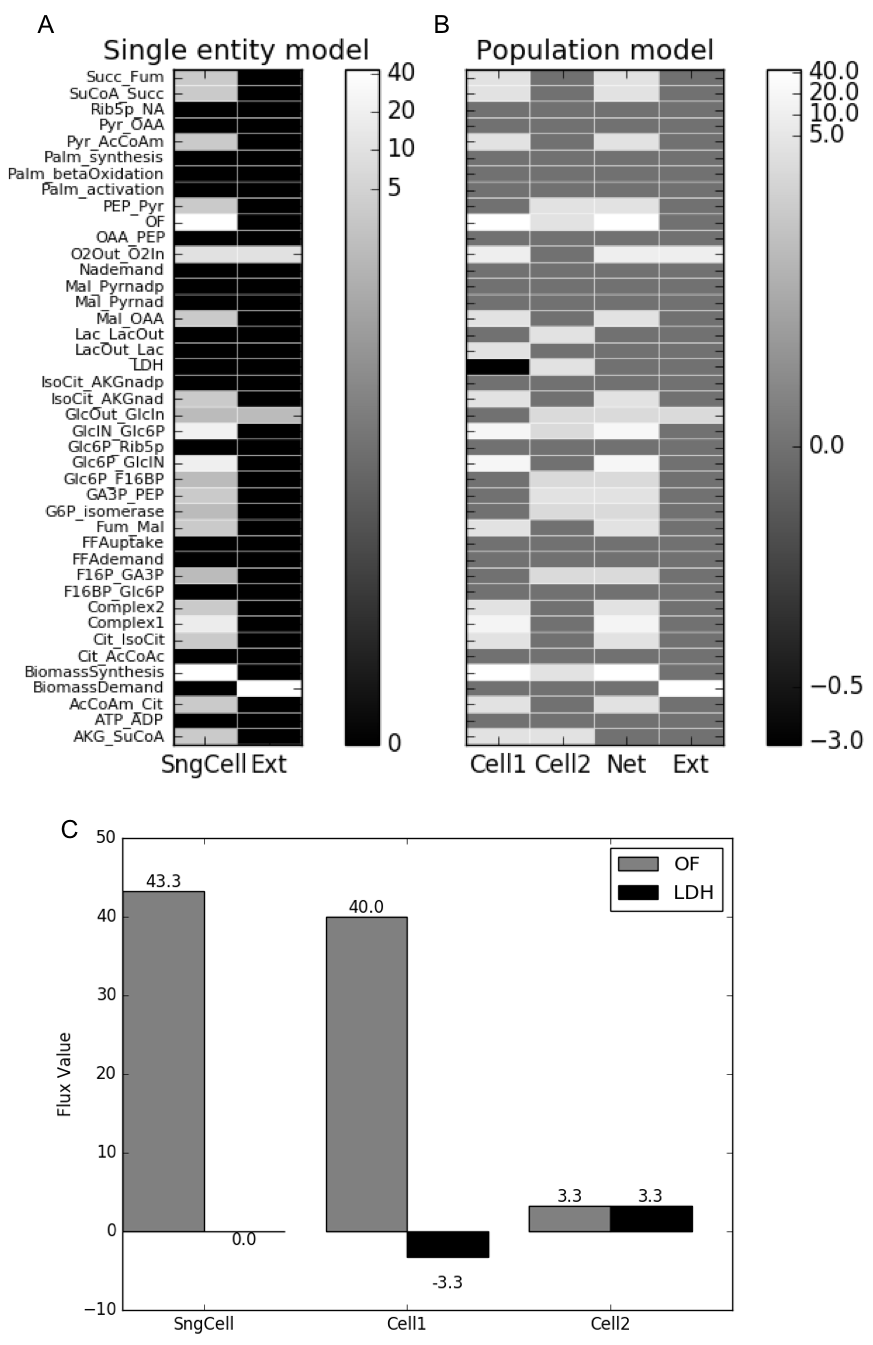}
\caption[Results obtained from the execution of Flux Balance Analysis on the single entity model and on the population model.]{Results obtained from the execution of Flux Balance Analysis on the single entity model and on the population model, by giving in both cases the same maximum amount of nutrients (glucose and oxygen). In all the heatmaps the color of each cell is proportional to the flux value of the corresponding reaction according to the gray chromatic scale on the right-hand side of each heatmap. Panel A, Heatmap showing the flux values for the reactions of the single entity model. The column SngCell contains the flux values of the internal reactions, whereas the column Ext contains the flux values of the exchange reactions. Panel B, Heatmap showing the flux values for the reactions of the population model. The column Cell1 contains the flux values of the internal reactions of the first identified subpopulation, the column Cell2 contains the flux values of the internal reactions of the second identified subpopulation, the column Net contains the net flux values of the internal reactions of the two different subpopulations, whereas the column Ext contains the flux values for the exchange reactions of the population model. This heatmap is useful to show that the net flux distribution of the multicomponents model perfectly mirrors the flux distribution obtained as a solution of the single entity model (panel A). Panel C, Bar plot showing the flux values for the objective function (referred to as ``OF'') and lactate production (referred to as ``LDH'') reactions in both the single entity and population models. The graph shows that following the maximization of ATP production, a heterogeneity at objective function flux value level emerged between the two subpopulations (columns ``Cell1'' and ``Cell2'') within the population model. The bar plot also shows that between the two interacting subpopulations, the one that is responsible for the secretion of lactate in the medium produces ATP at a lower rate compared to the subpopulation in which lactate is consumed} 
\label{fig:4}
\end{figure}

As a proof of concept of our methodology, we constructed a generic and non-compartmentalized toy-model, that we refer to as
``single entity core model'', based on the current knowledge on the metabolic pathways most involved in cancer metabolic rewiring (Fig. \ref{fig:2}). This model consists of 45 reactions and 40 metabolites and includes the following metabolic pathways: glycolysis, production and consumption of lactate, tricarboxylic acid cycle (TCA cycle), oxidative phosphorylation (OXPHOS), pentose phosphate pathway (PPP), palmitate synthesis and beta-oxidation of fatty acids. Uptake reactions for the nutrients glucose and oxygen have been added as constraint to the model for defining the cell medium composition, and the maximization of the ATP yield has been chosen as objective function of the system, as we are focused on the reprogramming of energy metabolism of cancer cells. \\
We used this toy-model as building block for constructing the ``population model'' characterized by the interaction among individual components, all having the same stoichiometry and sharing the same glucose and oxygen supply. As for the single entity model, we chose the maximization of the overall ATP production as objective function of the whole system. \\
We therefore investigated the potentialities of the constraint-based approach in the simulation of both the single entity and the population models in order to understand if this approach is able to highlight some differences between the two models in terms of their resulting flux distributions. The two models under investigation have the same objective function, equal exchange (sink and demand) reactions and the same boundaries on nutrients uptake.\\ 
We applied FBA to obtain the ATP production yield -- computed as the ratio between the objective function flux value and the number of entities included in the model - as a function of the simulated number of entities, including the classic case of one single entity.
We observed that the computed yield is constant (Fig. \ref{fig:3}) and, therefore, not affected by the number of entities. This outcome confirms that FBA on individual metabolic networks well approximates the average cell of an optimal population. In fact, the net flux distribution of the different cells perfectly mirrors the flux distribution obtained as a solution of the single FBA model (Fig. \ref{fig:4}, panels A and B). However, the population model allows to investigate the tumor population at a different level, elucidating the ways in which the average behavior can be achieved, how the individual cells may differ in their metabolism, and how different sub-populations of cells may interact with each other to attain the common goal.

\subsection{Metabolic heterogeneity within population models}\label{sec:ResultPart2}
\begin{figure}
\includegraphics[max height=15 cm,max width=12 cm]{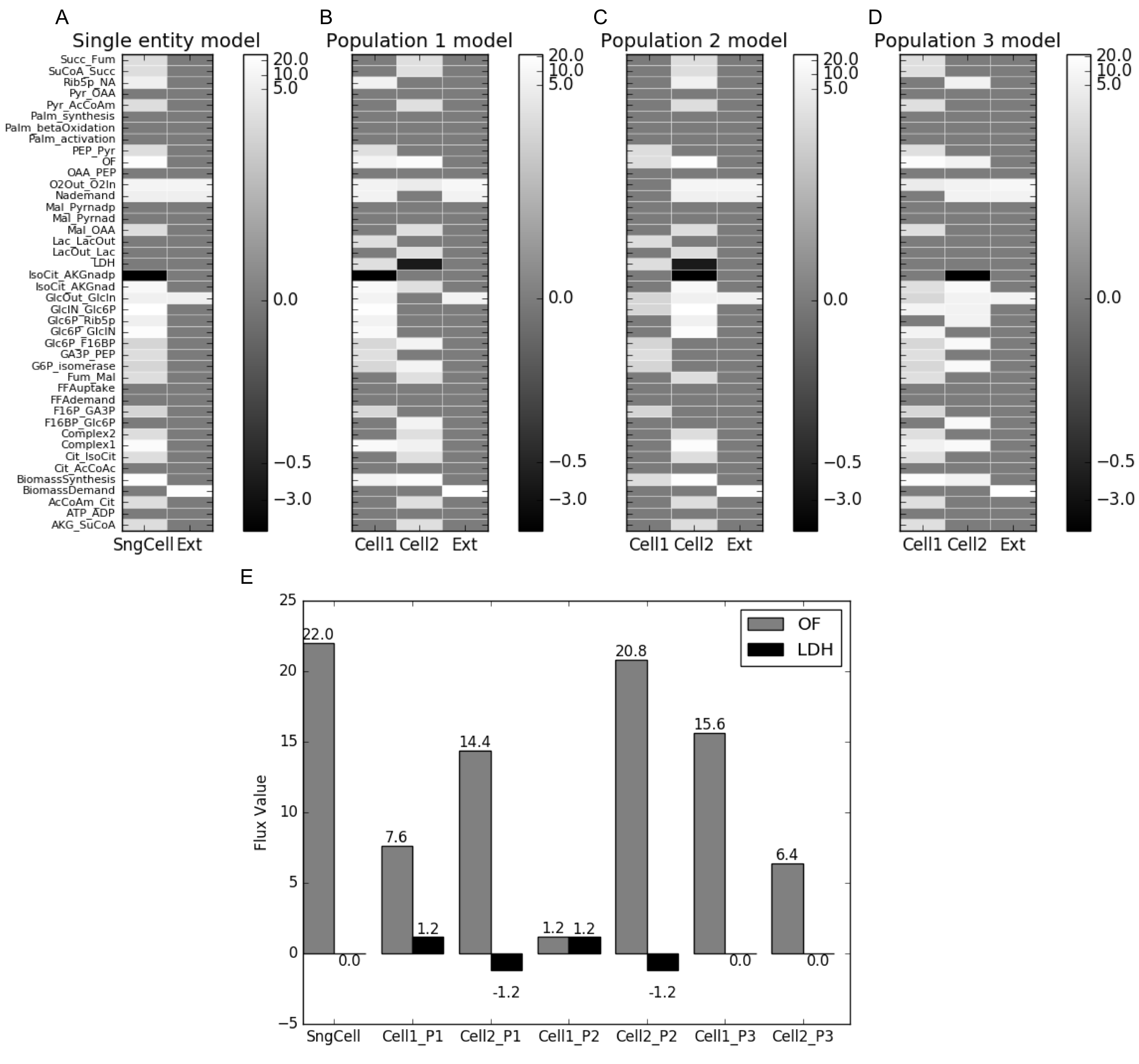}
\caption[Results obtained from the execution of Flux Balance Analysis on the single entity model and on the population model after a perturbation of the glucose/oxygen ratio.]{Results obtained from the execution of Flux Balance Analysis on the single entity model and on the population model after a perturbation of the glucose/oxygen ratio. In all the heatmaps the color of each cell is proportional to the flux value of the corresponding reaction according to the gray chromatic scale on the right-hand side of each heatmap. Panel A, Heatmap showing the flux values for the reactions of the single entity model. The column SngCell contains the flux values of the internal reactions, whereas the column Ext contains the flux values of the exchange reactions. Panels B-C-D, Heatmaps showing the three alternative and equally optimal flux distributions identified in the population model. The column Cell1 contains the flux values of the internal reactions of the first identified subpopulation, the column Cell2 contains the flux values of the internal reactions of the second identified subpopulation, whereas the column Ext contains the flux values for the exchange reactions of the model. Panel E, Bar plot showing the flux values for the objective function (referred to as ``OF'') and lactate production (referred to as ``LDH'') reactions in both the single entity and population models. The graph shows that following the maximization of ATP production, a heterogeneity at objective function flux value level emerged in the population model between the two subpopulations of each of the three identified alternative optimal populations (columns ``Cell1\_P1'' and ``Cell2\_P1'', columns ``Cell1\_P2'' and ``Cell2\_P2'', columns ``Cell1\_P3'' and ``Cell2\_P3''). The bar plot also shows, in all cases, that between the two interacting subpopulations of each population, the one that is responsible for the secretion of lactate in the medium produces ATP at a lower rate compared to the subpopulation in which lactate is consumed}
\label{fig:5} 
\end{figure}

We shifted the focus toward a more in-depth study of how the flux distribution identified in the single entity model distributes among multiple cells within the population model. We wanted to understand whether FBA approach could highlight the heterogeneity factor that we know to be a long-established knowledge of cancer populations, or, in alternative, if the different components belonging to the system just share out the common good. In other words, we tested if a cooperative behavior could arise within cancer population or if all tumor cells behave the same way for achieving the common goal, which is an enhanced and uncontrolled growth and proliferation. In this regard, we used the toy-model to generate a population model consisting of 10 components, which are assumed to be single cells that are representative of the metabolism of distinct subpopulations this cells belong to, all having the same topology and stoichiometry, and sharing the same glucose and oxygen supply. We performed FBA simulations on this system maximizing its overall ATP production and then we exploited FVA analysis in order to explore the variability range of the system across the alternative ways for obtaining the same objective function flux value. \\
Given the same maximum amount of glucose and oxygen to the system, the reached steady state is characterized by a particular ratio between glucose and oxygen uptake flux value of 1:6, which is known to be the correct ratio so that one glucose molecule is completely oxidize by oxygen. We observed that glucose uptake flux value is adjusted based on the quantity of oxygen that is available in the medium, and that all the entities constituting the population under investigation seek to maximize the common good for satisfying the common aim. This aspect, showed by the analysis of the flux distribution of the population model, emerged together with the observation that maximization of the ATP production by the population model is obtained following the interaction between two distinct subpopulations which show a very different ATP production rate and differ in their energy generating pathways (Fig. \ref{fig:4}, panels B and C). The first subpopulation, which probably corresponds to the hypoxic cancer cells, is composed by glucose-dependent cells that convert glucose into lactate that is then secreted in the medium; the second subpopulation, which probably corresponds to the better-ossigenated cancer cells, imports the lactate produced by the first subpopulation by using it as energy source instead of glucose, and is characterized by an active TCA cycle and respiratory chain. The flux distribution analysis showed that these two subpopulations do not contribute in an independent manner to the achievement of the common goal, but they cooperate with each other deriving mutual benefit from this interaction. \\
With changing environmental conditions as in Figure \ref{fig:5}, by perturbing the glucose to oxygen ratio and forcing the system towards more tumoral environmental conditions (i.e. constraining glucose uptake reaction flux to a higher levels than that we found previously), the system reached different steady states having in common the fact that an increasing glucose uptake corresponds to a lowering of the objective function value. This happens because both there is not enough oxygen so that glucose is completely oxidize, and we are in the case in which the individual can produce lactate whereas the entire population cannot. In addition to this result, we constantly noticed that, among the interacting subpopulations within the system, those ones that are responsible for the secretion of lactate in the medium, also produce ATP at a lower rate compared to the subpopulations in which lactate is consumed (Fig. \ref{fig:5}, panel E). The analysis of flux variability, through FVA, showed that there is not just one single possible way by which different components belonging to the population model can interact with each other. On the contrary, for the purpose of maximizing the chosen objective function, three different scenarios (Fig. \ref{fig:5}, panels B, C and D), which represent alternative optimal solutions, emerged. This outcome strenghtens the importance of the heterogeneity factor within cancer populations as a means that has been developed for evolutionary reasons to resist to anti-tumor treatments.

\section{Conclusions}
To investigate heterogeneity within cellular populations and as a complement to either single cell or standard metabolomics, we investigated the potentialities of a population model that is characterized by multiple interacting components, all having the same topology and stoichiometry, and sharing the same nutrients supply. These two elements were necessary for developing a methodology that would allow to identify within a population model which are the best strategies able to promote the cancer population growth and how many distinct subpopulations, characterized by different types of metabolism interact with each other within the same tumor population. The advantage of performing FBA simulations on a population model compared to that on single entity model is the possibility of identifying distinct subpopulations having different phenotypes, but coexisting within the same system, and the possibility of better understanding the heterogeneity degree within a cancer population. \\
Through our approach, we came to the conclusion that the entire cancer population can be represented, at a first level, through a single entity model which provides a snapshot of the average behavior of the cell population, and at a second level, through a network of metabolic networks, each of them representing the individual subpopulations. Indeed, just knowing the average behavior results in a limited outlook because the heterogeneity that might emerge within cancer population is not considered. Exploiting FBA method on a network consisting of multiple interacting components allowed us to observe that cancer cells cooperate with each other to reach a specific objective, and that they do not need to have the same metabolism type in order to reach the optimal value of objective function. Through our methodology we explored another level of complexity owned by cancer disease: the objective of the system does not correspond to the objectives of the individual entities since different subpopulations have different role within tumor tissue. \\
Since rewiring of energy generating pathways and enhanced growth are closely related, the results here obtained following the ATP production maximization, may be generalized to the case of maximization of biomass production in cancer population. Accordingly, we can say that also the main metabolic trait that unifies all cancer cells, which is an uncontrolled and enhanced proliferation, is not the common objective for all individual cells belonging to the system. According to the cancer stem cell theory, the tumor growth is not driven by all cells belonging to the cancer population, but it is mainly sustained by only a specific portion of the tumor that consists of the so-called cancer stem cells \cite{yoo2008,marusyk2010}. \\
Further analyses on increased complexity level metabolic models will be performed in order to further validate our methodology and to investigate whether the emergence of subpopulations which are characterized by both consumption of the lactate that is present within the medium, and a higher growth rate compared to the subpopulations that are responsible for the secretion of lactate in the medium, holds even for more biologically grounded and comprehensive metabolic models.\\
In conclusion, we would like to point out that, although our modeling approach may become computationally demanding when simulating a large number of genome-scale models, core models, that are limited to specific aspects of metabolism and require a higher level of abstraction, may be more effective in uncovering system-level properties of cancer metabolic rewiring and the interpretation of the outcomes of their simulation is significantly more straightforward than that of genome-scale networks \cite{difilippo2016}.\\
Furthermore, we would also like to emphasize that the approach discussed here, is not tailored to just analysing cancer cells populations, but it may be suitable for exploring, in general, how the interactions among more than one component (such as different types of healthy cells, bacteria, yeasts) may influence the overall behavior of a population for which a mismatch between the objective of the individual members and that of the entire population is assumed.

\section*{Acknowledgement}

The institutional financial support to SYSBIO Center of Systems Biology - within the Italian Roadmap for ESFRI Research Infrastructures - is gratefully acknowledged. C.D., R.C. and M.D. are supported by SYSBIO fellowships.

\bibliography{bibliography}

\end{document}